\documentclass[pra]{revtex4}
\usepackage[dvips]{graphicx}
\usepackage{graphicx,epsfig}
\usepackage{amsmath,amsfonts,amssymb,amsthm,amsbsy,mathtools,mathrsfs}
\usepackage{bbold}

\newcommand{\ket}[1]{\mathop{\left|#1\right>}\nolimits}            
\newcommand{\bra}[1]{\mathop{\left<#1\,\right|}\nolimits}         

\newcommand{\brk}[2]{\langle #1 | #2 \rangle}
\newcommand{\kbr}[2]{| #1\rangle\!\langle #2 |}

\newcommand{\oxb}{\overset{\sim}{\otimes}}                

\newcommand{\Tr}[1]{\mathop{{\mathrm{Tr}}_{#1}}}          

\def\dg{\dagger}

\def\vp{\varphi}

\begin{document}

\author{Kamil Br\'adler}
\email{kbradler@cs.mcgill.ca}
\affiliation{
    School of Computer Science,
    McGill University,
    Montreal, Quebec, H3A 2A7, Canada
    }
\author{Roc\'io J\'auregui}
\affiliation{
    Instituto de F\'isica, Universidad Nacional Aut\'onoma de M\'exico, Apdo. Postal 20-364, M\'exico D.F. 01000
    }

\title{Comment on ``Fermionic entanglement ambiguity in noninertial frames''}

\begin{abstract}
In this comment we show that the ambiguity of entropic quantities calculated in~\cite{unruh-fermi} for fermionic fields in the context of Unruh effect is not related to the properties of anticommuting fields, as claimed in~\cite{unruh-fermi}, but rather to wrong mathematical manipulations with them and not taking into account a fundamental superselection rule of quantum field theory.
\end{abstract}

\maketitle

The notion of entanglement and how it is defined for fermionic systems is known to be a subtle issue~\cite{fermions}. On the mathematical side, it is the concept of braiding that turns out to be crucial~\cite{braiding}. On the physical side, a superselection rule that forbids superpositions of fermion states with different parities. We argue that by not considering both crucial concepts the authors of~\cite{unruh-fermi} are led to an incorrect conclusion that there is an ambiguity in the amount of entanglement in multi-partite fermion states depending on the operator ordering. In order to support our claim we start by presenting the minimal amount of information needed to study systems of fermions~\footnote{Note that in this comment when we talk about particles or modes we mean the same thing. We are allowed to do so since every mode contains at most one particle (fermion).}.

Fermion annihilation and creation operators ($a,a^\dg$) satisfy the canonical anticommutation relations that are also known as the CAR algebra
\begin{equation}\label{eq:CAR}
    [a,a^\dg]_+=1,
\end{equation}
where $[,]_+$ stands for the anticommutator. Different modes (fermions) will be labeled by different letters ($a,b$ and $c$) and they hide the two relevant fermion degrees of freedom (the momentum and spin). The intermode relations read
\begin{equation}\label{eq:interCAR}
    [a,b]_+=[a,b^\dg]_+=[a^\dg,b^\dg]_+=0
\end{equation}
and similarly for the $c$ mode. Therefore, when we act on a vacuum with $a^\dg b^\dg$, the order of the operators matters
\begin{equation}\label{eq:order}
    a^\dg b^\dg\ket{vac}=\ket{11}_{ab}\neq\ket{11}_{ba}.
\end{equation}
We conclude that
\begin{equation}\label{eq:notensor}
    \ket{11}_{ab}\neq\ket{1}_a\otimes\ket{1}_b,
\end{equation}
where $\otimes$ denotes the standard tensor product. This is an obvious statement -- only in the case of two qubits or two distinguishable particles the order is irrelevant and $\ket{11}_{ab}=\ket{1}_a\otimes\ket{1}_b=\ket{1}_b\otimes\ket{1}_a$ holds in this case.

To fully account for the CAR algebra we have to introduce the concept of braiding. This structure comes from the theory of quantum groups~\cite{braiding} and the main physical application today probably lies  in the physics of anyons~\cite{anyons} and topological quantum field theory. A braided (or twisted) tensor product
\begin{equation}\label{eq:braidedtensor}
    \ket{1}_a\oxb\ket{1}_b=e^{i\vp}\ket{1}_b\oxb\ket{1}_a
\end{equation}
describes the whole family of statistics which, in a certain sense, interpolates between bosons and fermions. As a matter of fact, fermions are the simplest non-trivial example of the braided statistics where $\exp{i\vp}=-1$.
Hence, we have to understand Eq.~(\ref{eq:notensor}) in the following way:
\begin{equation}\label{eq:twistedtensor}
    a^\dg b^\dg\ket{vac}\equiv a^\dg\oxb b^\dg\ket{vac}=\ket{1}_a\oxb\ket{1}_b.
\end{equation}
To avoid clumsy notation in the rest of the text we will understand the braided tensor product for $\vp=\pi$ and abbreviate the RHS of Eq.~(\ref{eq:twistedtensor}) as $\ket{11}_{ab}$. But we will keep in mind that it is not an ordinary tensor product.

The adjoint operator for fermions is well defined and acts as
\begin{equation}\label{eq:daggeraction}
\dg:a^\dg\overset{\sim}{\otimes} b^\dg\mapsto -a\overset{\sim}{\otimes}b.
\end{equation}
Eq.~(\ref{eq:daggeraction}) necessarily follows from the anticommutation relations for particles obeying the fermion statistics and the normalization condition for fermion multi-particle  states:
\begin{equation}\label{eq:adjointactingonVAC}
    -\bra{vac}\big(a\oxb b\big)\big(a^\dg\oxb b^\dg\big)\ket{vac}
    =-\bra{11}_{ab}\ket{11}_{ab}=\bra{11}_{ba}\ket{11}_{ab}=1.
\end{equation}

Having all the necessary mathematical concepts ready we may discuss the issues with~\cite{unruh-fermi}. There are two incorrect statements on page two in~\cite{unruh-fermi}. First of all, the authors claim in Eq. (3) that the two-fermion state 
$$
\ket{\Psi}={1\over2}(\ket{00}+\ket{01}+\ket{10}+\ket{11})_{ab}
$$
can be factorized as
\begin{equation}\label{eq:wrongfactorization}
\ket{\Psi}={1\over2}\Big[\ket{0}_a+\ket{1}_a\Big]\otimes\Big[\ket{0}_b+\ket{1}_b\Big],
\end{equation}
where $\otimes$ stands for an ordinary tensor product. The states $\{\ket{0},\ket{1}\}$ are suddenly treated by the authors as qubits. But this is not permitted -- fermions do not become qubits just by changing their name. These two structures are algebraically very different. One could argue to consider the states in Eq.~(\ref{eq:wrongfactorization}) to be fermions and substitute the tensor product with the correct braided tensor product. But even this is not possible. There is another, this time physical, reason why Eq. (3)~in~\cite{unruh-fermi} does not represent a valid state. We will discuss it shortly.

The second algebraic error committed on page~two of~\cite{unruh-fermi} will be highlighted on their Eq. (5). The authors claim that by changing the order of particles in a three-particle fermionic state followed by tracing over one particle one can get different results for different orderings. This is exemplified on the following state (\cite{unruh-fermi}, Eq. (5))
\begin{equation}\label{eq:threeparticlestate}
    \ket{\Phi}={1\over2}(\ket{100}+\ket{010}+\ket{101}+\ket{011})_{abc},
\end{equation}
where we explicitly indicate the mode labels.  As argued above, the Hermitian conjugation is a braided operator~\cite{braiding} and so for fermionic fields it acts as
\begin{equation}\label{eq:threeparticlestate_bra}
    \dg:\ket{\Phi}\mapsto\bra{\Phi}={1\over2}(\bra{100}+\bra{010}-\bra{101}-\bra{011})_{abc}.
\end{equation}
We check the norm and find it positive $\brk{\Phi}{\Phi}=1$. Note that the adjoint operator acts on $\ket{\Phi}$ in the sense of Eq.~(\ref{eq:daggeraction}). To trace over a subsystem (for instance, the particle occupying the $c$-mode) we follow the rules of the CAR algebra and we obtain
\begin{equation}\label{eq:traceover}
    \Tr{c}{\kbr{\Phi}{\Phi}}={1\over2}(\ket{10}+\ket{01})_{ab}(\bra{10}+\bra{01})_{ab}.
\end{equation}
Following~\cite{unruh-fermi}, we change the order of the modes $c$ and $b$ in~(\ref{eq:threeparticlestate}) and get
\begin{equation}\label{eq:threeparticlestate_swapped}
    \ket{\Phi'}={1\over2}(\ket{100}+\ket{001}+\ket{110}-\ket{011})_{acb}.
\end{equation}
By tracing over the $c$ mode in $\kbr{\Phi'}{\Phi'}$ we have to consistently use the CAR algebra. In this case, when we `skip' the occupied modes $a$ or $b$ by $\ket{1}_c$ or $\bra{1}_c$ the sign has to change. As a consequence, we again end up with state~(\ref{eq:traceover}) contrary to the observation made in~\cite{unruh-fermi} (a separable state Eq.~(6)). A different result would be a sign of inconsistency of the whole CAR algebra.

As we indicated the mathematical aspects of fermionic systems are not the only issue in~\cite{unruh-fermi}. The states on which the authors demonstrate various inconsistencies violate a boson-fermion superselection rule. Going back, for instance, to the fermionic state
$$
\ket{\Psi}={1\over2}(\ket{00}+\ket{01}+\ket{10}+\ket{11})
$$
(Eq.~(3),~\cite{unruh-fermi}) or the above analyzed state $\Phi$, we see that the authors did not take into account the fact that the Fock space for fermions is a direct sum of two subspaces containing even and odd parity fermion states and their superpositions cannot be formed. We may say that this is a special case of the superselection rule for bosons and fermions which is commonly discussed in the quantum field theory literature~\cite{QFT}.

In passing we make a few observations. In the rest of their article, the authors draw a number of conclusions based on assumptions we have shown to be incorrect. It is not clear how the rest of the article is affected. One  conclusion from~\cite{unruh-fermi} is particularly perplexing. The authors found that in their Refs.~[1-6,8-11,20,22], where fermionic systems and their classical and quantum correlations were studied in the context of Unruh effect, many of the results do not agree with one another (when calculating comparable quantities). It seems that if the braiding properties of fermions and the superselection rule were taken properly into account and a large number of results in those articles should be reconsidered. An explicit example of such correction can be found in~\cite{longversion} for the entanglement of formation. It was demonstrated that this entanglement measure cannot be used for fermions as it has been defined. This is another subtle consequence of the superselection rule. We point out that, to our knowledge, the only work on fermionic Unruh effect where the braiding properties of fermions were considered is~\cite{grass}. Certain entropic quantities used in quantum Shannon theory were proved to remain valid but only for a restricted class of fermionic states very different from the ones used in~\cite{unruh-fermi} and the references within.

\begin{acknowledgements}
K. B. would like to thank Prakash Panangaden and Tomas Jochym-O'Connor for discussions. This work was supported by a grant from the Office of Naval Research (N000140811249).
\end{acknowledgements}

\end{document}